**Charged cell membrane in electrolyte**

M. Pekker[1], M. N. Shneider[2], M. Keidar[1]

[1]The George Washington University, Northwest Washington, DC 20052
[2]Department of Mechanical and Aerospace Engineering, Princeton University, Princeton NJ 08544

Corresponding Author: M.Pekker: pekkerm@gmail.com, Ph: 512-293-7785

**Abstract**

An effect of membrane surface charge on mechanical properties of the phospholipid bilayer membrane and pores formation is considered. It is shown that the outer and inner surfaces of the phospholipid membrane is always subject to tension, regardless of magnitude and sign of the charges on its surface. This is due to the fact that the Debye length of the extracellular and intracellular electrolyte is always much smaller than the thickness of the membrane, and accordingly the electric field on the outer surface of the membrane is always much larger than the field inside the membrane. This result contradicts the generally accepted notion that a charged phospholipid membrane is similar to a charged capacitor in a dielectric medium and it is always subject to compression. Phospholipid bilayer membrane will be subject to compression only in a very weak electrolyte (~1mM/L), when Debye length is larger than the membrane thickness. It is also shown that the membrane surface charges lead to pore compression when the pore radius $R_p$ is larger than the Debye radius $r_D$ and to the stretching when $R_p < r_D$. Difference in the coefficients of surface and edge tension of phospholipid cell membranes can be explained by taking into account the cell membrane surface charge. Simple experiments are proposed to test the influence of the cell membrane surface charge on its mechanical properties.

**Introduction**

Great number of experimental and theoretical work has been devoted to investigation of properties of cell membranes and penetration of ions through the membrane, see, for example, monographs [1-3]. Also many studies have been focused on ion transport through the cell membrane pore (see, for example, [4-6]). To that end molecular dynamics approaches were used to model ion transport across the phospholipid membrane [7] and formation of pores on the phospholipid membranes located in the deionized water [8].

The goal of this paper is to draw an attention to the role of conductivity of a solution in living organisms on the mechanical properties of the phospholipid bilayer membrane and to the formation of pores in it. All our calculations and estimates are made within the framework of the phenomenological theories developed in [9-11], in which the phospholipid membrane of a cell is considered as a thin shell, described by the coefficients of surface and edge tension taken from the known experiments. Note, that dependence of the coefficient of the surface tension of the phospholipid membrane on the conductivity of the solution was also found experimentally [12-14]. The results obtained in this work are of a qualitative nature and amenable for experimental verification.

It is known that the surface of phospholipid bilayer membrane immersed into the electrolyte is charged negatively, since the heads of the phospholipids are always facing the positive charge outward (see, for example, [14]). This fact was used in Refs [15, 16] to estimate the surface charge of the phospholipid membrane of cells. It was also shown, that ions motion along the membrane is restricted due to strong electrostatic interactions ions with phospholipid dipolar heads [15,16]. In fact it was found that the ion binding energy is about several $k_B T$ (where $T$ is the temperature, $k_B$ is Boltzmann constant. Note that the distribution of ions near the surface of a phospholipid membrane was calculated using molecular dynamics [17,18] without considering the binding energy.

In the first part of this article we consider an effect of membrane surface charge on mechanical properties of the phospholipid bilayer membrane pore formation. It is shown that the outer and inner surfaces of the phospholipid membrane are always subject to tension, regardless of the magnitude and sign of the charges on its surfaces. This is due to the fact that the Debye length of the extracellular and intracellular electrolyte is always much smaller than the thickness of the membrane, and accordingly the electric field on the outer surface of the membrane is always much larger than the field inside the membrane Fig.1.

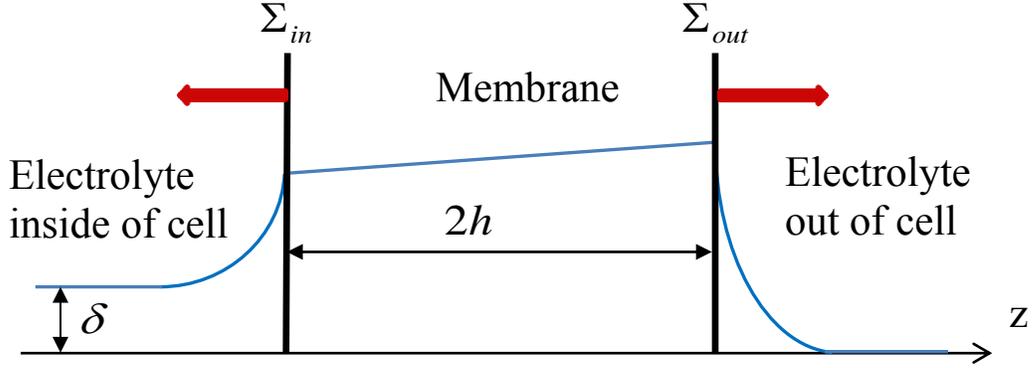

Fig. 1. *Schematics of potential distribution across the membrane. Debye length is much smaller than the membrane thickness. The arrows show the directions of the forces acting on surface charges $\Sigma_{in}$ and $\Sigma_{out}$ ($\Sigma_{in}, \Sigma_{out} > 0$). These forces are determined by potential drop in electrolytes outside and inside the cell. $\delta$ is potential inside of cell.*

Expressions for forces, shown in Fig. 1, acting on the inner and outer sides of the membrane have the form:

$$F_{in} = E_{in}\Sigma_{in} < 0, \qquad E_{in} = -\left.\frac{\partial\varphi}{\partial z}\right|_{-h+0} - \left.\frac{\partial\varphi}{\partial z}\right|_{-h-0} \approx -\left.\frac{\partial\varphi}{\partial z}\right|_{-h+0} < 0 \qquad (1)$$

$$F_{out} = E_{out}\Sigma_{out} > 0, \qquad E_{out} = -\left.\frac{\partial\varphi}{\partial z}\right|_{h+0} - \left.\frac{\partial\varphi}{\partial z}\right|_{h-0} \approx -\left.\frac{\partial\varphi}{\partial z}\right|_{h+0} > 0 \qquad (2)$$

Note that $F_{in}$ it is always negative and $F_{out}$ – positive , regardless of the signs $\Sigma_{in}, \Sigma_{out}$ (see (8) and (9)) .

This result contradicts the generally accepted opinion that a charged phospholipid membrane is similar to a charged capacitor in a dielectric medium, is always subject to compression [9-11]. Phospholipid bilayer membrane will be subject to compression only in a very weak electrolyte, when the ion density $n_i \approx 1 [\text{mM/L}]$, that is, almost three orders of magnitude less than the ion density into extracellular and intracellular liquids of living organisms. It has to be mention, that in the weak electrolytes membrane charge density is proportional to the ion density in electrolytes while in the strong electrolytes membrane charge density cannot exceed surface density of phospholipids in the membrane [15,16]. Thus charge density of the membrane in DI water should be zero.

The rest of the paper is organized as follows. In Section 1 it is shown that the increase in the thickness of the double phospholipid membrane with increasing solubility of the solution, observed in [19], can be explained by an increase in the surface charge density on the membrane. The same mechanism also makes it possible to explain the membrane thickness increase [20,21]. and axon radius, when the action potential passes [20,22-30].

In part 2 it is shown that taking into account the surface charge on the membrane makes it possible to explain the difference in the coefficients of surface and edge tension of phospholipid cell membranes [9,11].

In sections 3 and 4 we will consider effect of membrane surface charge on formation of pores. In these sections it will be shown that if the pore radius $R_p$ is greater than the Debye length $r_D$, the surface charge prevents growth of pores, in opposite $R_p < r_D$, the collaps of pores is suppresed.

1. **Charged phospholipid bilayer membrane of cell in electrolyte**

Typical intracellular and extracellular ion densities are shown in Table 1 [31]. As an example let us consider muscle fibers having ion composition close to that of neurons [31]. In this case Debye length is about $r_D = 0.75 \cdot 10^{-9}$ nm, while it is about $r_D = 0.43 \cdot 10^{-9}$ nm $r_D = 0.43 \cdot 10^{-9}$ nm for squid axon [31]. In all considered cases $r_D << d = 2h << R$, where $d$ is the membrane thickness ( $5\text{nm} \leq d \leq 10\text{nm}$), and $R$ is the characteristic size of cell. As such we can consider membrane as a plane surface as shown schematically in Fig. 2.

**Table.1**: Intracellular and extracellular ion densities (following [31]).

| Ions | Squid Axon | | muscle fibers (neurons) | |
|---|---|---|---|---|
| | Intercellular mM/L | extracellular mM/L | Intercellular mM/L | extracellular mM/L |
| $Na^+$ | 50 | 460 | 10 | 125 |
| $Ka^+$ | 400 | 10 | 124 | 2 |
| $Ca^{2++}$ | – | 10 | 5 | 2 |
| $Mg^{2+}$ | 10 | 54 | 14 | 1 |
| total | 460 | 534 | 153 | 130 |
| | | | | |
| $Cl^-$ | 135 | 560 | 2 | 77 |
| $HCO_3^-$ | – | – | 12 | 27 |
| $(A)^-$ | 345 | – | 74 | 13 |
| Others | – | – | 65 | 13 |
| Total | 460 | 560 | 153 | 130 |

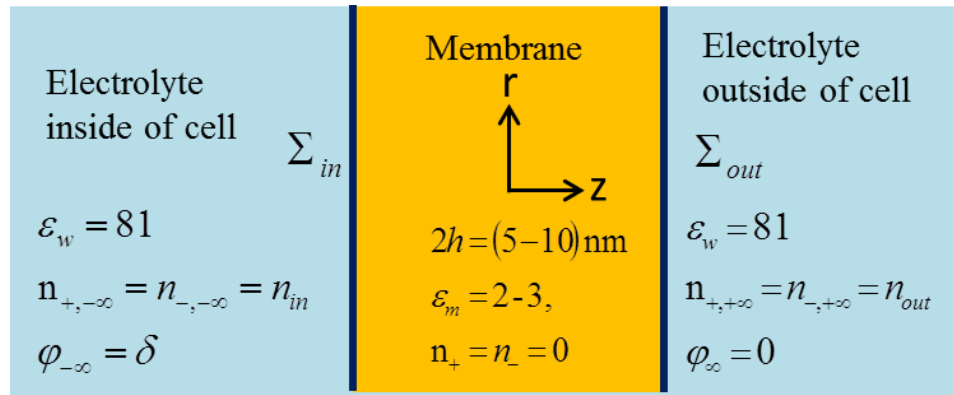

Fig.2. *Model of phospholipid bilayer membrane in electrolyte. $\Sigma_{in}$, $\Sigma_{out}$ are surface charges on the inner and outer surface of membrane respectively, $\varepsilon_m$, $\varepsilon_w$, – dielectric permittivity in membrane and electrolyte corespondently. $n_{in}$, $n_{out}$ – ion densities inside of cell and outside far from membrane. Charge density inside the membrane is zero.*

In this case, the distribution of potential in the membrane is described by the Laplace equation, and outside of the membrane by the Poisson equation (Appendix A).

Since the potential difference, $\delta$ (see Fig.1), between the electrolyte inside and outside of the cell varies from 0 to -0.09 V [32], the surface charge on the membrane lies within $\Sigma = 0.3 - 0.002$ C/m$^2$ [32,33], the half-thickness of the membrane $h$ varies from 2.5 to 5 nm, one can argue that the contribution of $\delta$ at the calculation of the potentials and fields can be neglected (see A6-A7) at:

$$\Sigma > \frac{\delta \varepsilon_0 \varepsilon_m}{2h} \left( \frac{2h \varepsilon_w}{r_D \varepsilon_m} \right)^{1/2} \approx 4 \cdot 10^{-3} [\text{C/m}^2] \tag{3}$$

In (3) we used $h = 5$ nm, $\varepsilon_m = 2.5$, and $\delta = 0.09$.

With these parameters we have (see A6,A8,A11,A12):

$$\varphi_{out} \approx \frac{r_D \Sigma_{out}}{\varepsilon_w \varepsilon_0}, \quad \varphi_{in} \approx \frac{r_D \Sigma_{in}}{\varepsilon_w \varepsilon_0}, \tag{4}$$

$$E_{h+0} \approx \frac{\Sigma_{out}}{\varepsilon_w \varepsilon_0}, \quad |E_{h+0}| >> |E_{h-0}|, \tag{5}$$

$$E_{-h-0} \approx -\frac{\Sigma_{in}}{\varepsilon_w \varepsilon_0}, \quad |E_{-h-0}| >> |E_{-h+0}|, \tag{6}$$

$$W \approx \frac{r_D}{4\varepsilon_0 \varepsilon_w} \left( \Sigma_{in}^2 + \Sigma_{out}^2 \right) \tag{7}$$

That is, the energy associated with the presence of a charged phospholipid membrane (7) is mainly determined by the charge on it, that is, by the density and composition of ions in the electrolyte. The forces acting on the unit area of the outer and inner surfaces of the membrane are equal, respectively:

$$F_{out} = E_{h+0} \Sigma_{out} \approx \frac{\Sigma_{out}^2}{\varepsilon_w \varepsilon_0}, \tag{8}$$

$$F_{in} = E_{-h-0} \approx -\frac{\Sigma_{in}^2}{\varepsilon_w \varepsilon_0}, \tag{9}$$

It can be seen that $F_{in}$ always directed into the cell, and $F_{out}$ – out the cell, as it shown on Fig.1. Subtracting (9) from (8) and, assuming $\Sigma_{out} = \Sigma_{in} = \Sigma$, we obtain the value of the wedging force acting on the cell membrane:

$$F_{\Sigma,streach} = F_{out} - F_{in} = \frac{\Sigma_{out}^2}{\varepsilon_w \varepsilon_0} + \frac{\Sigma_{in}^2}{\varepsilon_w \varepsilon_0} = 2\frac{\Sigma^2}{\varepsilon_w \varepsilon_0} \tag{10}$$

It should be pointed out that tension force does not depend on the sign of charge on inner and outer surface of membrane. $F_{E,streach}$ leads to increase of the cell membrane thickness. Calculated force (according to eq. 10) is shown in Table 2:

**Table 2:** Dependence of tension pressure on the membrane surface charge in the case of $\Sigma_{out} = \Sigma_{in} = \Sigma$.

| $\Sigma [C/m^2]$ | $F_{\Sigma,streach}[kPa]$ |
|---|---|
| 0.002 | 11.2 |
| 0.004 | 44.8 |
| 0.008 | 179.2 |
| 0.016 | 716.8 |
| 0.032 | 2867.2 |

It was shown in [19] that the thickness of the double phospholipid membrane depends on the density of ions in the solution, which corresponds to an increase in the wedge forces when the surface ion density on the surface of the membrane changes. It should be pointed out that according to experiments [20, 21] the propagation of the action potential is associated with the increase of the membrane thickness. This effect can be explained by the changes in the surface charge on the membrane, and, correspondingly, by the increase in the wedging forces predicted by the current model.

Total pressure acting on the membrane can be calculated as follows:

$$F_{\Sigma,total} = F_{out} + F_{in} = \frac{1}{\varepsilon_w \varepsilon_0}\left(\Sigma_{out}^2 - \Sigma_{in}^2\right) \tag{13}$$

Let us turn attention to Eq. 13. The main negative ion in extracellular electrolyte is $Cl^-$, while intracellular ions are mainly aspartic acid anion $AspA^-$, glutamic acid anion $GluA^-$ and bicarbonate ion $HCO_3^-$ (Table 1). Since $Cl^-$ is much smaller than intracellular ions and according to [15,16] one can assume that $|\Sigma_{out}| > |\Sigma_{in}|$ and $F_{\Sigma,total} > 0$.

A large number of experimental studies have shown that the action potential propagation is accompanied by fast and temporary mechanical changes. These changes include an axonal radius [22–29] and pressure [24,30]. According to this model a rapid change in the surface density of ions $\Sigma_{out}$ and $\Sigma_{in}$ associated with the ejection of ions as the action potential passes, leads to the appearance of uncompensated force $F_\Sigma$ (13), which may alter the radius of the axon observed in Refs [22-29].

In the weak electrolyte absolute values of electric fields out of membrane are always much less than inside, therefore ( see A11):

$$\Sigma_{out} = -\Sigma_{in} = \Sigma, \quad \delta = \frac{2h\Sigma_{in}}{\varepsilon_0 \varepsilon_m}, \quad E_m = \frac{\delta}{2h} = -\frac{\Sigma}{\varepsilon_0 \varepsilon_m} \tag{11}$$

and

$$F_\Sigma = E_m \Sigma_{out} = -\frac{\Sigma^2}{\varepsilon_0 \varepsilon_m}, \quad W = \frac{\Sigma^2 h}{\varepsilon_0 \varepsilon_m} \tag{12}$$

Note that usually the cell membrane is considered as a capacitor in a non-conducting medium (see, for example, [9-11]), and, accordingly, the charge on it leads to compression of the membrane with a force per unit surface equal to $F_{\Sigma,m}$.

## 2. Forces acting on the surface of the pore in the phospholipid membrane

According the conventional wisdom (see, for example, [34, 35]), the pore in the phospholipid membrane has the form shown in Fig.3. Tensile forces determine the curvature of the surface at the boundary of the pore, and constricting forces control the hole in the membrane.

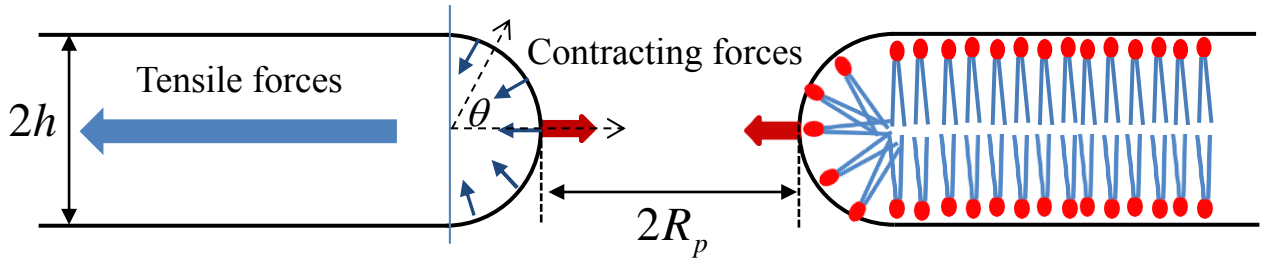

Fig.3. *Schematic view of the pore in the membrane (the heads are directed into the pores). Arrows show the forces acting on the surface of the pores.*

Tensile forces can be calculates as:

$$F_t = \int_{-\pi/2}^{\pi/2} \frac{\sigma}{h} 2\pi \left( R_p + h(1-\cos(\theta)) \right) h \cos(\theta) d\theta = 4\pi\sigma R_p + 4\pi\sigma h - \pi^2 \sigma h, \tag{13}$$

while contracting forces are:

$$F_c = -\int_{-\pi/2}^{\pi/2} \frac{\sigma}{R_p + h(1-\cos(\theta))} 2\pi \left( R_p + h(1-\cos(\theta)) \right) h \cos(\theta) d\theta = 4\pi\sigma h \tag{14}$$

In (22) and (23) $\sigma$ is the coefficient of surface tension of the membrane, $h$ is the half of the thickness of the membrane, $R_p$ is the radius of the pore. According to [11], $\sigma$ lies within 2 to 4 mN/m, depending on the salinity of the electrolyte. Since intracellular and extracellular pH lies in the range 5.5 - 8, it follows from [36], that $\sigma = 2.34$ mN/m. Adding (22) and (23), we obtain:

$$F_\sigma = F_t + F_c = 4\pi\sigma R_p - \pi^2 \sigma h \qquad (15)$$

It follows from (15) that if the critical radius of the pore, at which it grows is equal to $R_{cr} = \pi h/4$, and independent on the surface tension $\sigma$.

For example, at $h = 2.5nm$, the critical size at which the pore irreversibly grows is equal to $R_{cr} \approx 2$nm. Substituting the values $\sigma = 2.34$ mN / m and $h = 2.5nm$ in Eq. (22), we obtain:

$$F_\sigma = 2.96 \cdot 10^{-2} R_p - 5.8 \cdot 10^{-11} [N] \qquad (16)$$

The following expression for the forces acting on the pore surface for the phospholipid membrane with $h = 2.5nm$ is given in [11, 37], obtained on the basis of experimental data [10, 13], without taking into account the forces associated with the surface charge:

$$F_\sigma^* = \gamma R_p - \Gamma = 6.28 \cdot 10^{-3} R_p - 5.8 \cdot 10^{-11} \text{ [N]} \qquad (17)$$

Accordingly, crirical pore radius can be estimated, $R_{cr} \approx 10$ nm. Note that in the next section, we show that such a strong difference in the first term of Eqs. (16) and (17) can be explained by the surface charge on the phospholipid membrane located in the electrolyte. Radius of the pore $R_p$ in charged membrane immersed in electrolyte is always larger than the Debye radius $r_D$, $R_p > r_D$.

### 3. Forces acting on the surface of the pore in the phospholipid membrane immersed in electrolyte

Let us consider forces acting on a pore of a charged membrane shown schematically in Fig. 4. We consider a case with surface charge on inner and outer surface being the same i.e. $\Sigma_{out} = \Sigma_{in} = \Sigma$. Note that the surface charge prevents pore closing if pore size is smaller than the Debay length. On the other hand if pore size is larger than the Debay length the cell membrane surface charge restrict pore growth. Recall that electrostatic repulsion of edges of the small pore is similar to effect of growth of a small hole in the electrolyte [34,35].

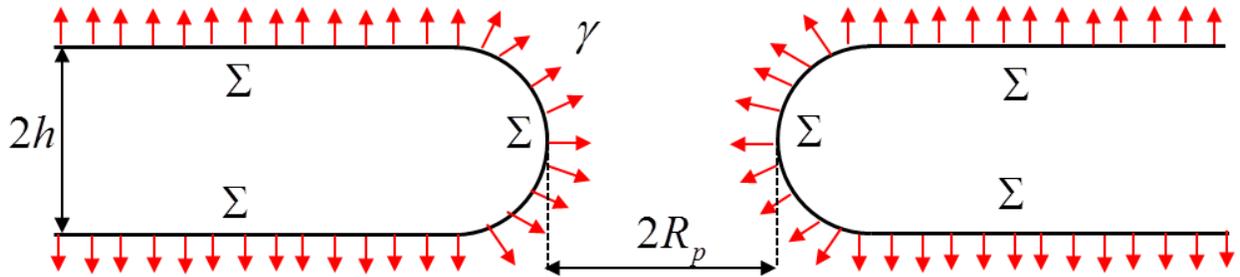

Fig. 4. *Pore in the cell membrane $R_p > r_D$, $\Sigma$ and is the surface charge on the inner and outer surfaces of the membrane. The forces acting on the pore are shown by arrows.*

Let us find the compressive force acting on the pore related to the surface charge under the assumption $R_p > r_D$. Integrating (8) over the surface of the pore surface, we obtain:

$$F_\Sigma = -\int_{-\pi/2}^{\pi/2} \frac{\Sigma^2}{\varepsilon_w \varepsilon_0} 2\pi(R_p + h(1-\cos(\theta)))h\cos(\theta)d\theta = -(4\pi R_p + (4-\pi)h)\frac{h\Sigma^2}{\varepsilon_w \varepsilon_0} \quad (18)$$

Recall that the expression (18) is valid if $R_p > r_D \approx 0.7$ nm.

Taking into account (15), the force acting on the pore is equal to

$$F_{total} = F_t + F_c + F_\Sigma = 4\pi R_p\left(\sigma - \frac{h\Sigma^2}{\varepsilon_w \varepsilon_0}\right) - \left(\pi^2 \sigma + (4-\pi)\frac{h\Sigma^2}{\varepsilon_w \varepsilon_0}\right)h \quad (19)$$

Substituting the values $\sigma = 2.34$ mN/m and $h = 2.5$ nm into (19) and equating the obtained expression $F_\sigma^*$, we obtain

$$\Sigma_* \approx 0.023 \, [\text{C/m}^2] \quad (20)$$

The obtained value of $\Sigma_*$ is reasonable, since, as noted above, the surface charge lies within $0.3 - 0.002 \, [\text{C/m}^2]$. It follows from (19), that the surface charge of the membrane at which pores can not exist, is equal to:

$$\Sigma_0 = \sqrt{\frac{\varepsilon_w \varepsilon_0 \sigma}{h}} \quad (21)$$

At $\sigma = 2.34$ mN/m and $h = 2.5$ nm, we obtain $\Sigma_0 = 0.0257 \, [\text{C/m}^2]$. Once again, we note that all our estimates are valid for the condition $R_p > r_D$.

From (19) we find that the critical radius at which the pore will grow can be estimated as

$$R_{cr} = \frac{\pi^2 \sigma + (4-\pi)\dfrac{h\Sigma^2}{\varepsilon_w \varepsilon_0}}{4\pi\left(\sigma - \dfrac{h\Sigma^2}{\varepsilon_w \varepsilon_0}\right)} h \quad (22)$$

The expression for the work required to create a pore of the radius $R_p$:

$$A = -\int_0^{R_p} F_{total} dR = -2\pi R_p^2\left(\sigma - h\frac{\Sigma^2}{\varepsilon_w \varepsilon_0}\right) + R_p h\left(\pi^2 \sigma + (4-\pi)h\frac{\Sigma^2}{\varepsilon_w \varepsilon_0}\right) \quad (23)$$

We note that the energy associated with the change in the pore size (accriding to Eq. 7):

$$W_{p,s} \approx \pi R_p^2 W = \frac{\pi R_p^2 r_D \Sigma^2}{2\varepsilon_0 \varepsilon_w} \tag{24}$$

This energy is much less than the work that should be done against the forces $F_\Sigma$ (see (23)):

$$W_\Sigma \approx \frac{2\pi R_p^2 h \Sigma^2}{\varepsilon_w \varepsilon_0} \tag{25}$$

Let us now consider pore in charged membrane which is immersed in electrolyte. In order to calculate the electrostatic forces acting on a pore let us consider a simplified model shown schematically in Fig. 5.

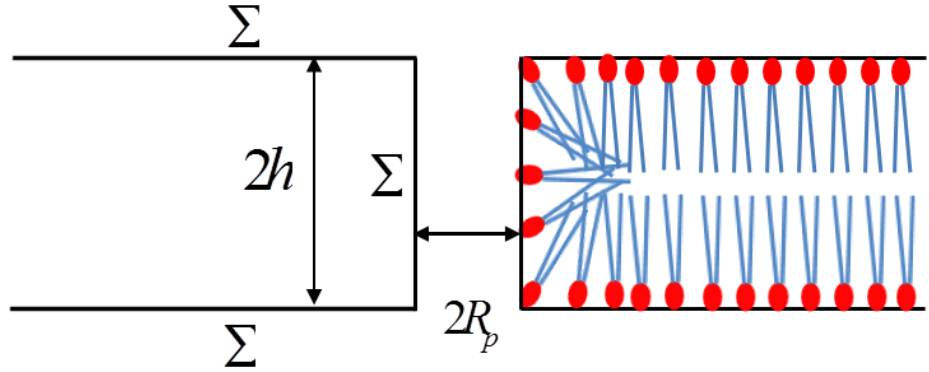

Fig. 5 *Model of the pore in the cell membrane.*

For simplicity, we assume that $R \ll h$. In this case we can neglect edge effects and use 1D Poisson equation to calculate the potential distribution inside the pore, Appendix B.

Dependence of the pressure $P_\Sigma^*$ (B12) on dimensionless pore radius $\xi_p = R_p / r_D$ is shown in Fig. 6. When $R_p > 2r_D$, pressure $P_\Sigma^*$ only weakly depends on radius of the pore. It is easy to see that to within the accuracy of $(4-\pi)h/4\pi R_p$, the value $4\pi h R_p \tilde{P}_\Sigma$ coincides with the force $F_\Sigma$ in (18).

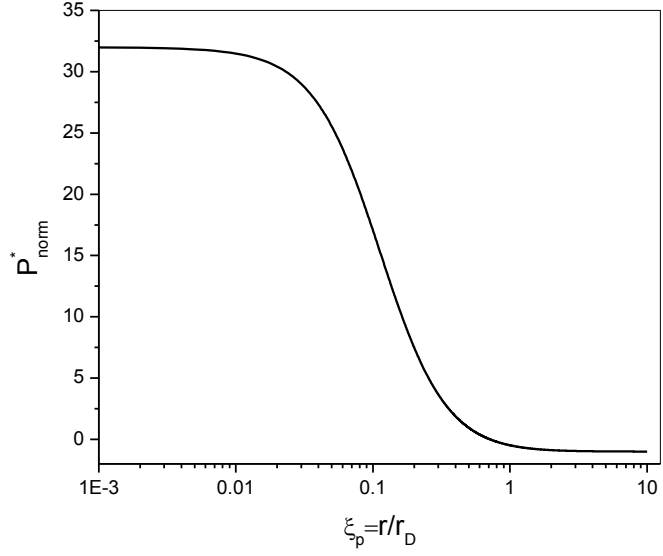

**Fig.6.** *The dependence* $P^*_{norm} = P^*_\Sigma / \tilde{P}_\Sigma$ *on* $\xi_p = R_p / r_D$.

If $\xi_p \ll 1$ (see B12):

$$\frac{P^*_\Sigma}{\tilde{P}_\Sigma} \approx \frac{\varepsilon_w}{\varepsilon_m}, \qquad \tilde{P}_\Sigma = \frac{\Sigma^2}{\varepsilon_0 \varepsilon_w} \tag{26}$$

i.e. change on the membrane prevents pore closing.

On the other hand if $\xi_p \gg 1$ $I_1(\xi_p) \approx I_1(\xi_p)$ and thus:

$$\frac{P^*_\Sigma}{\tilde{P}_\Sigma} \approx -1, \tag{27}$$

i.e. charge on the membrane leads to pore closing.

The question of the influence of the external field on the formation of pores in the membrane and their development requires special consideration and goes beyond the scope of this article.

### 4. On the critical size of the pore

In this Section we calculate the critical pore size of the phospholipid membrane.

Recall that the following formula was used in [9-11] for the energy of the pore:

$$W_p(R_p) = 2\pi\gamma hR - \pi R_p^2 \Gamma - \pi R_p^2 h \left( \frac{\varepsilon_w - \varepsilon_m}{\varepsilon_m \varepsilon_w} \right) \frac{\Sigma^2_{out}}{\varepsilon_0} \tag{28}$$

Where $\Gamma \approx 10^{-3}$ N/m, $h = 5$ nm, $\varepsilon = 2.5$, $\gamma \approx 2 \cdot 10^{-3}$ N/m.

Within this framework the pore formation can be described as shown schematically in Fig.7. The third term in (28) corresponds to the change in the energy of the capacitor in a nonconducting medium (see (12)). However such simple analogy might not be correct. This is due to the following two arguments: (1) in the dionized water the surface charge on the membrane equals zero, therefore third term in (28) has to be canceled out, and (2) in a strong electrolyte, energy of capacitor is given by formula (7).

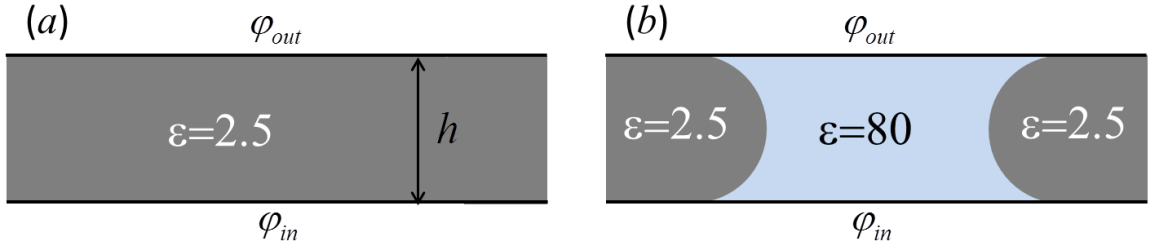

Fig.7. *The illustration of a pore formation. (a) – section of the membrane before rupture. (b) – membrane section after the rupture (pore formation).*

The dependencies of the energy necessary to create the pores $W_p = -A_p$ (see (23) and the the force acting on its surface $F_p = \partial W_p / \partial R_p$ on the radius of the pore for different values of the surface charge $\Sigma$ at $h = 2.5$ nm, $\sigma = 2.43$ mN/m are shown in Fig. 8

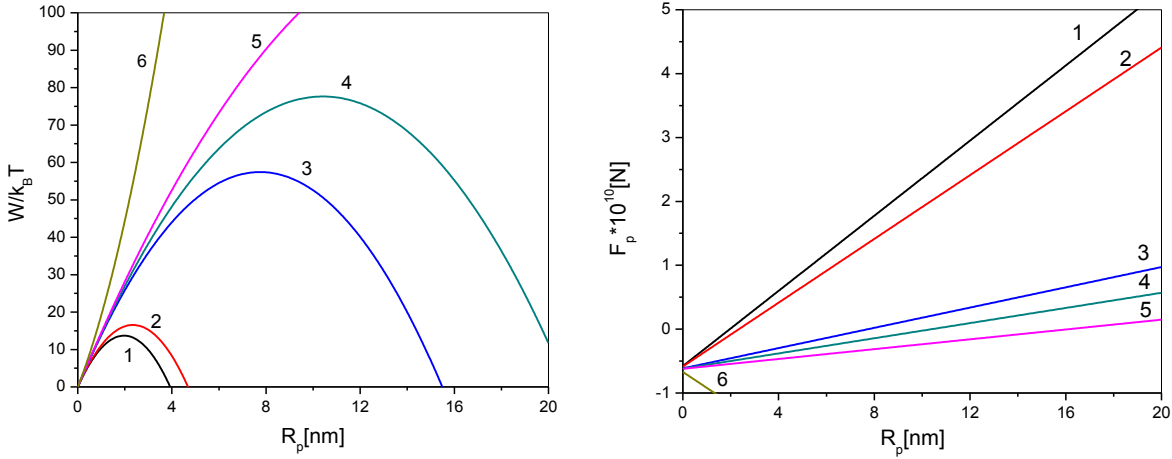

Fig.8. *The dependencies of $W_p = -A_p$ (see (23) and $F_p = -\partial A_p / \partial R_p$ on the radius of the pore for different values of the surface charge $\Sigma$. Line 1 corresponds $\Sigma = 0$, 2 – $\Sigma = 0.01\,C/m^2$, 3 – $\Sigma = 0.022$ $C/m^2$, 4 – $\Sigma = 0.023\,C/m^2$, 5 – $\Sigma = 0.024\,C/m^2$, 6 – $\Sigma = 0.035\,\,C/m^2$. At $\Sigma = 0.035$ pore will not be formed.*

Figure 9 shows the dependence of the critical radius of the pores on the charge on it (22) ат $h = 2.5$ nm, $\sigma = 2.43$ mN/m.

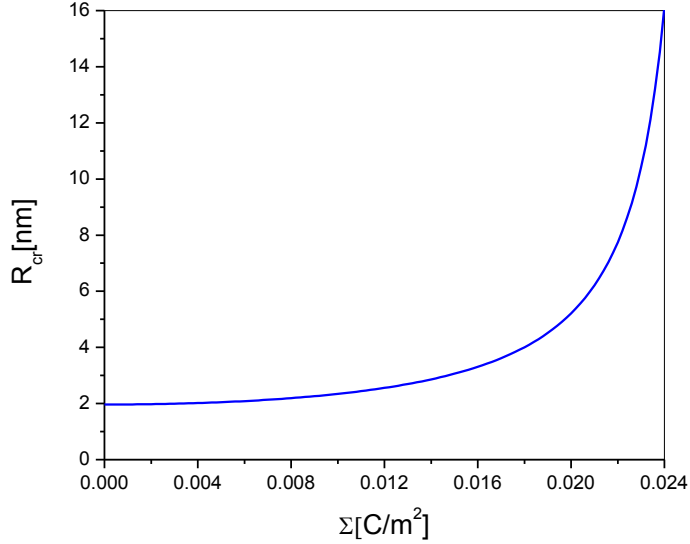

Fig. 9. *The dependence of the critical radius of the pore on the surface charge.*

Note that the above analysis is correct when number of ions in a pore is large. Let us evaluate this assumption. According to Fig. 4 volume of a pore can be calculated as $V = \pi h(R^2 + h^2/12)$. Considering the ion density of about $n_\infty = 10^{26} m^{-3}$ (that corresponds to 150 mMol/L), the minimum number of ions in pore space is about $N = Vn_\infty = \pi h^3 n_\infty /12 = (3.14 \cdot 10^{-24}/12) \cdot 10^{26} = 25$. In the case of a critical pore radius of about 10 nm, number of ions in the pore space is about 300.

The mechanism described in this paper can be also related to relatively large pores responsible for transport of large molecules and ions such as water and $H_2O_2$ [38,39]. To this end surface charge and its modification in case of interaction of cell with plasma jet is important.

Recall that first term in right hand side of Eq. 31-33 asymptotically approaches zero as pore radius tend to zero. As such results described above are valid for pore radius $R > h$. Dynamics of small pores requires special consideration and goes above frameworks on this paper.

It should be pointed out that this model can be used to predict membrane changes as results of the plasma treatment. This is of particular interest for adaptive plasma platform [40]. To this end the present model can be implemented in predictive model argorithm as a part of adaptive approach.

Let us outline some possible experiments that can validate this model preductions. This includes measurements several key properties of the membrane such as mebrane thickness, pore size and ion transport. The disjoining forces acting on the double phospholipid membrane depend on the density of the surface charge on it, which in turn depends on the ion density in the electrolyte. As such changing the composition of the electrolyte and the density and composition of the ions in it, one can measure the dependence of the thickness of the double phospholipid membrane from the strength and type of ions in the electrolyte. In addition we have shown that presence of a surface charge on the membrane reduces the probability of nanometer pores formation. Therefore, as the density of the electrolyte ions increases, suppression of the molecular diffusion through the nanometer-sized membrane might be observed. The surface charge of the membrane depends on the type of ions . According to the model prediction the membrane should bend towards the lighter negative ions.

**Conclusions:**
1. It has been demonstrated that mechanical and physiological properties of the cell membrane are determined by the membrane surface charge. Membrane surface charge depends on the ion composition in intracellular and extracellular environment. It is predicted that changing ion composition of extracellular electrolyte can control pore formation and transport properties of the cell membrane. Taking into account that healthy and non-healthy cells have different ion composition it is possible to develop selective treatment by controlling ion composition of extracellular medium.
2. It is shown that the thickness increase of the bilayer phospholipid membrane at the increasing of the salinity of the solution, observed in [19], can be explained by the increasing of the surface charge density on the membrane. The same mechanism also makes it possible to explain the membrane thickness increasing [20,21], as well as the axon radius increasing at the action potential propagation [20,22-30].
3. It is shown that taking into account the surface charge on the membrane makes it possible to explain the difference in the coefficients of surface and edge tension of phospholipid cell membranes.
4. It is shown, that if the pore radius $R_p$ is greater than the Debye radius $r_D$, the surface charge prevents growth of pores, in opposite case, i.e. $R_p < r_D$, the surface charge prevents the compression of pores.

**Acknowledgements**
This work was supported in part by GW Institute of Nanotechnology and National Science Foundation, grants: 1465061 and PHY-1463867.

**APPENDIX A**

Let us consider a system of governing equations for potential distribution for the configuration shown in Fig 2. To this end we assume that the potential at the membrane boundary is small and ions are singly charged:

$$\frac{e\varphi_{out}}{2k_B T}, \frac{e\varphi_{in}}{2k_B T} < 1, \tag{A1}$$

Where e is the electron charge and $T = 300$ K is the temperature of electrolyte.

In this case, the problem of the potential distribution in the membrane is described by the Laplace equation, and outside – by the Poisson equation:

$$\frac{d^2\varphi}{dz^2} = \begin{cases} -\dfrac{\varphi}{r_{D,out}}, & z > h \\ 0, & -h \leq z \leq h \\ -\dfrac{\varphi}{r_{D,in}}, & z \leq -h \end{cases}, \tag{A2}$$

With the boundary conditions on $\pm\infty$

$$\varphi|_{\infty} = 0, \quad \varphi|_{-\infty} = \delta \tag{A3}$$

and at the boundaries of the membrane:

$$\varepsilon_0\varepsilon_m \frac{d\varphi}{dz}\bigg|_{h-0} - \varepsilon_0\varepsilon_w \frac{d\varphi}{dz}\bigg|_{h+0} = \Sigma_{out}, \quad \varepsilon_0\varepsilon_w \frac{d\varphi}{dz}\bigg|_{-h-0} - \varepsilon_0\varepsilon_m \frac{d\varphi}{dz}\bigg|_{-h+0} = \Sigma_{in} \tag{A4}$$

In (A2), $r_{D,in}$, $r_{D,out}$ are the Debye radii in the electrolyte outside and inside the cell. Since the density of ions outside of the cell differs from the density of the ions inside by no more than 20% (see, Table 1), we will further assume that $n_{in} = n_{out} = n_{\infty}$, and, accordingly, $r_{D,in} = r_{D,out} = r_D = \left(\varepsilon_w\varepsilon_0 k_B T / 2e^2 n_{\infty}\right)^{1/2}$.

Recall that according to Refs [15,16] the surface density of charges depends not only on the density of ions in the solution, but also on the type of ions. Therefore, at the same density of ions outside and inside the cell, the surface charge on the inner and outer surface of the membrane can vary greatly.

Solution of system of equations (A2) with boundary conditions (A3) и (A4) has the form:

$$\varphi = \begin{cases} \varphi_{out} e^{-(z-h)/r_{Dt}} & z \geq h \\ \varphi_0 + \dfrac{z}{2h}\varphi_1 & -h \leq z \leq h \\ \varphi_{in} e^{(z+h)/r_D} + \delta & z \leq -h \end{cases} \tag{A5}$$

where:

$$\varphi_{out} = \frac{r_D}{\varepsilon_w\varepsilon_0}\left(\Sigma_{out} - \frac{\dfrac{\varepsilon_m r_D}{2h\varepsilon_w}(\Sigma_{out} - \Sigma_{in}) - \delta\dfrac{\varepsilon_0\varepsilon_m}{2h}}{\left(1 + \dfrac{\varepsilon_m r_D}{\varepsilon_w h}\right)}\right) \approx \frac{r_D}{\varepsilon_w\varepsilon_0}\left(\Sigma_{out} + \delta\frac{\varepsilon_0\varepsilon_m}{2h}\right) \tag{A6}$$

$$\varphi_{in} = \frac{r_D}{\varepsilon_0\varepsilon_w}\left(\Sigma_{in} + \frac{\dfrac{\varepsilon_m r_D}{2h\varepsilon_w}(\Sigma_{out} - \Sigma_{in}) - \delta\dfrac{\varepsilon_0\varepsilon}{2h}}{\left(1 + \dfrac{\varepsilon r_D}{\varepsilon_w h}\right)}\right) \approx \frac{r_D}{\varepsilon_0\varepsilon_w}\left(\Sigma_{in} - \delta\frac{\varepsilon_0\varepsilon_m}{2h}\right) \tag{A7}$$

$$\varphi_1 = \frac{(\Sigma_{out} - \Sigma_{in})}{1 + \dfrac{r_D\varepsilon_m}{h\varepsilon_w}}\frac{r_D}{\varepsilon_0\varepsilon_w} - \frac{\delta}{1 + \dfrac{r_D\varepsilon_m}{h\varepsilon_w}} \approx (\Sigma_{out} - \Sigma_{in})\frac{r_D}{\varepsilon_0\varepsilon_w} - \delta \tag{A8}$$

$$\varphi_0 = \frac{r_D}{\varepsilon_0\varepsilon_w}(\Sigma_{out} + \Sigma_{in}) + \delta \tag{A9}$$

In (A6-A8) we took into account that $\frac{er_D}{2\varepsilon_w h} = \frac{3 \cdot 0.75 nm}{81 \cdot 10 nm} = 2 \cdot 10^{-3}$ is a small parameter of the problem[1].

After substitute (A5) and (A6) into (A1) one can find constrain on charge surface density:

$$\Sigma_{in}, \Sigma_{out} < 2\varepsilon_0 \varepsilon_w k_B T / er_D \approx 0.05 \, C/m^2. \tag{A10}$$

The electric field corresponding to (A5) has the form:

$$E = -\frac{\partial \varphi}{\partial z} = \begin{cases} \frac{1}{\varepsilon_w \varepsilon_0}\left(\Sigma_{out} + \delta \frac{\varepsilon_0 \varepsilon_m}{2h}\right) e^{-(z-h)/r_D} & z \geq h \\ -\frac{1}{2h}\left((\Sigma_{out} - \Sigma_{in})\frac{r_D}{\varepsilon_0 \varepsilon_w} - \delta\right) & -h \leq z \leq h \\ -\frac{1}{\varepsilon_0 \varepsilon_w}\left(\Sigma_{in} - \delta \frac{\varepsilon_0 \varepsilon_m}{2h}\right) e^{(z+h)/r_D} & z \leq -h \end{cases} \tag{A11}$$

The expression for the energy associated with the charged membrane:

$$W = \frac{1}{2}\int_{-\infty}^{\infty} E^2 \varepsilon dz = \frac{r_D}{4\varepsilon_0 \varepsilon_w}\left(\Sigma_{in}^2 + \Sigma_{out}^2 + \delta^2 \frac{\varepsilon_0^2 \varepsilon_m^2}{2h^2} - \delta \frac{\varepsilon_0 \varepsilon_m}{h}(\Sigma_{out} - \Sigma_{in})\right) + \\ \frac{r_D}{4\varepsilon_0 \varepsilon_w h} \cdot \frac{\varepsilon_m}{\varepsilon_w} \cdot \frac{r_D}{2h}(\Sigma_{out} - \Sigma_{in})^2 + \frac{\varepsilon_0 \varepsilon_m}{4h}\delta^2 - \frac{\varepsilon_m}{\varepsilon_w} \cdot \frac{r_D}{2h}(\Sigma_{out} - \Sigma_{in})\delta \tag{A12}$$

**APPENDIX B**

To this end let us consider for simplicity that $R << h$. In this case we can neglect edge effects and use 1D Poisson equation to calculate the potential distribution inside the pore:

$$\frac{d^2\varphi}{d\xi^2} + \xi \frac{d\varphi}{d\xi} - \xi^2 \varphi = 0, \quad \xi = \frac{r}{r_D} \tag{B1}$$

The solution of this equation has the following form:
$$\varphi = A \cdot I_0(\xi), \quad \xi \leq \xi_p = R_p / r_D \tag{B2}$$

Corresponding electric field equal to:
$$E = -\frac{\partial \varphi}{\partial r} = -A \frac{I_1(\xi)}{r_D} \tag{B3}$$

Inside the membrane i.e. $\xi \geq \xi_p$ potential $\varphi_m$ is determined by Laplace equation:

---

[1] The solution of equations (A2) with the boundary conditions (A3) and (A4) can be found here:
http://erg.biophys.msu.ru/wordpress/wp-content/uploads/2013/04/bilayer-adsorption-task.pdf

$$\frac{1}{r}\frac{d}{dr}\left(r\frac{d\varphi_m}{dr}\right) = 0 \tag{B4}$$

with solution:

$$\varphi_m = B \cdot \ln(\xi) + C, \quad E_m = -\frac{\partial \varphi_m}{\partial r} = -\frac{B}{r_D \xi} \tag{B5}$$

Assuming that for $\xi = \xi_1 = \xi_p + h/r_D$ the potential $\varphi_m$ equals to the membrane potential in the absence of pore:

$$B\ln(\xi_1) + C = \frac{r_D \Sigma}{\varepsilon_0 \varepsilon_w}, \tag{B6}$$

Substituting $C$ from (B6) into (B5) yields:

$$\varphi_m = B\ln(\xi/\xi_1) + \frac{r_D \Sigma}{\varepsilon_0 \varepsilon_w} \tag{B7}$$

from the continuity condition for potential and electric field at the pore boundary one can obtain:

$$AI_0(\xi_p) = B\ln\left(\frac{\xi_p}{\xi_1}\right) + \frac{r_D \Sigma}{\varepsilon_0 \varepsilon_w} \tag{B8}$$

$$I_1(\xi_p) - \frac{\varepsilon_m}{\xi_w \xi_p} B = \frac{r_D \Sigma}{\varepsilon_0 \varepsilon_w} \tag{B9}$$

From (B8) and (B9) we can find unknown constants:

$$B = \frac{I_0(\xi_p) - I_1(\xi_p)}{I_1(\xi_p)\ln(\xi_p/\xi_1) - I_0(\xi_p)\varepsilon_m/(\xi_p \varepsilon_w)} \cdot \frac{r_D \Sigma}{\varepsilon_0 \varepsilon_m} \tag{B10}$$

$$A = \frac{\ln(\xi_p/\xi_1) - \varepsilon_m/(\xi_p \varepsilon_w)}{I_1(\xi_p)\ln(\xi_p/\xi_1) - I_0(\xi_p)\varepsilon_m/(\xi_p \varepsilon_w)} \cdot \frac{r_D \Sigma}{\varepsilon_0 \varepsilon_m} \tag{B11}$$

From known electric field one can calculate the pressure on the membrane as a product of the total electric field and the surface charge:

$$P_\Sigma^* = (E(\xi_p) + E_m(\xi_p))\Sigma = \frac{I_0(\xi_p) - \left(\xi_p \ln\left(\frac{\xi_1}{\xi_p}\right) + \left(1 + \frac{\varepsilon_m}{\varepsilon_w}\right)\right)I_1(\xi_p)}{I_1(\xi_p)\xi_R \ln\left(\frac{\xi_1}{\xi_p}\right) + I_0(\xi_p)\frac{\varepsilon_m}{\varepsilon_w}} \cdot \frac{\Sigma^2}{\varepsilon_0 \varepsilon_w} \tag{B12}$$